\documentclass[conference]{IEEEtran}
\IEEEoverridecommandlockouts
\usepackage{cite}
\usepackage{amsmath,amssymb,amsfonts}
\usepackage{algorithmic}
\usepackage{graphicx}
\usepackage{textcomp}
\usepackage{xcolor}
\usepackage{fancyhdr} 
\newcommand{\RomanNumeralCaps}[1]
    {\MakeUppercase{\romannumeral #1}}
\def\BibTeX{{\rm B\kern-.05em{\sc i\kern-.025em b}\kern-.08em
    T\kern-.1667em\lower.7ex\hbox{E}\kern-.125emX}}

\begin{document}

\newcommand{\highl}[1]{{\color{black} #1}}

\title{A CNN Approach for 5G mmWave Positioning Using Beamformed CSI Measurements}


\author{\IEEEauthorblockN{Ghazaleh Kia}
\IEEEauthorblockA{\textit{Department of Computer Science} \\
\textit{University of Helsinki}\\
Helsinki, Finland \\
ghazaleh.kia@helsinki.fi}
\and
\IEEEauthorblockN{Laura Ruotsalainen}
\IEEEauthorblockA{\textit{Department of Computer Science} \\
\textit{University of Helsinki}\\
Helsinki, Finland \\
laura.ruotsalainen@helsinki.fi}
\and
\IEEEauthorblockN{Jukka Talvitie}
\IEEEauthorblockA{\textit{Unit of Electrical Engineering} \\
\textit{Tampere University}\\
Tampere, Finland \\
jukka.talvitie@tuni.fi}

}


\maketitle

\begin{abstract}
The advent of Artificial Intelligence (AI) has impacted all aspects of human life. One of the concrete examples of AI impact is visible in radio positioning. In this article, for the first time we utilize the power of AI by training a Convolutional Neural Network (CNN) using 5G New Radio (NR) fingerprints consisting of beamformed Channel State Information (CSI). By observing CSI, it is possible to characterize the multipath channel between the transmitter and the receiver, and thus provide a good source of spatiotemporal data to find the position of a User Equipment (UE). We collect ray-tracing-based 5G NR CSI from an urban area. The CSI data of the signals from one Base Station (BS) is collected at the reference points with known positions to train a CNN. We evaluate our work by testing: a) the robustness of the trained network for estimating the positions for the new measurements on the same reference points and b) the accuracy of the CNN-based position estimation while the UE is on points other than the reference points. The results prove that our trained network for a specific urban environment can estimate the UE position with a minimum mean error of $0.98 \ m$. 

\end{abstract}

\begin{IEEEkeywords}
5G New Radio (NR), Artificial Intelligence (AI), Channel State Information (CSI), Convolutional Neural Network (CNN), Fingerprinting, Machine Learning (ML), Outdoor Positioning
\end{IEEEkeywords}

\section{Introduction}
In urban areas, where Global Navigation Satellite Systems (GNSS) signals are only partially available and degraded by multipath, it is difficult to reach a satisfactory positioning performance for many important use cases. To address the difficulty and achieve an accurate position solution, expensive and professional sensors such as Inertial Measurement Units (IMUs), cameras or other Radio Frequency (RF) signals are required to assist the positioning solution. 

Radio signals enable the position estimation by different methods, such as fingerprinting and signal features \cite{klus2,5g-deep}, range measurement and multilateration \cite{rets}, and the combination of the ranging and Angle of Arrival (AoA) \cite{aoa}. In addition, known or estimated channel properties of signals which are known as Channel State Information (CSI) are able to characterize the environment by describing how a signal propagates from a transmitter to a receiver. 
Consequently, in dense urban areas the multi-path effect resulted from the environment, is embedded in the CSI data. Thus, CSI is a good candidate for positioning purposes, where one of the use cases is fingerprinting \cite{csi2,loss}. Fingerprinting is mainly made of two phases: offline and online. In the offline phase, a large data set of fingerprints for each known position is collected and fed to the system. Later in the online phase, the test data is compared with the data set and maps the collected data to the corresponding position \cite{g-book}. Compared to traditional ranging and angular based positioning methods, where Non-Line-of-Sight (NLOS) propagation can considerably decrease the accuracy \cite{ION}, fingerprinting can exploit the richer features in NLOS scenarios to obtain a better positioning solution. 


CSI fingerprints of \highl{Wi-Fi} systems have been frequently used for indoor positioning \cite{csi-wifi,loss,wifi_fp}. Nevertheless, in outdoor urban areas Wi-Fi signals are not always available, or they are of poor quality. On the other hand, in the emerging 5G networks, efficient communications require continuous exchange of CSI data between the BS and the User Equipment (UE) \cite{complex}, which makes 5G New Radio (NR) CSI a convenient fingerprint in the urban area. 

The new millimeter Wave (mmWave) frequency band, introduced in the 5G NR, can be used to reach sub-meter level positioning accuracy \cite{wym1}. However, due to the high frequencies, the mmWave signals suffer from high penetration and path loss as well as the rain fading, which typically limits the number of available BSs for positioning. On the other hand, the short wavelength in millimeter level allows implementation of larger antenna arrays compared to the previous technologies \cite{wym1}. Partly resulting from the increased array sizes, another supplement provided by 5G NR technology is the beam-based operation, which enables highly directional beamforming via Massive Multiple Input Multiple Output (MIMO) technology \cite{beamforming}. Therefore, in this work, the utilized CSI data is beamformed and the CSI observations are performed per beam. These beam-wise CSI measurements work as an input to the proposed Convolutional Neural Network (CNN) based positioning engine. 


\begin{table*}[ht]
\caption{Neural Network-based Fingerprinting Methods}
\begin{center}
\begin{tabular}{|c|c|c|c|c|}

 \hline
\textbf{Reference} & \textbf{Environment} & \textbf{Radio Signal} & \textbf{Fingerprint} & \textbf{\highl{Minimum Mean Error}} \\ 
 \hline
\cite{csi2} & indoor & Wi-Fi & CSI & 1.8 m \\
\hline
\cite{csi-wifi2} & indoor & Wi-Fi & RTT & 0.91 m \\
\hline
\cite{loss} & indoor & Wi-Fi & CSI & 1.16 m \\
\hline
\cite{klus2} & outdoor & 5G and GNSS & radio signal strength & 1.75 m \\
\hline
\cite{5g-deep} & outdoor & 5G & RSRP and beam IDs & 1.4 m \\
\hline
\cite{5g_csi} & indoor & 5G & CSI from 4 BSs & 1.17 m\\
\hline
Proposed Work & outdoor & 5G & CSI from 1 BSs with beamforming technology & 0.98 m \\
 \hline

\end{tabular}
\label{fp}
\end{center}
\end{table*}

The main contributions of this paper are as follows:
\begin{enumerate}
    \item We introduce a novel utilization of 5G mmWave in position estimation. To the best of our knowledge, 5G NR CSI-based fingerprinting utilizing beamforming technology has not been reported in the literature.
    \item For obtaining the results, we utilize a realistic ray-tracing-based CSI data, considering a real-life city area with rich multipath propagation while following the guidelines of 5G NR specifications\cite{3GPPchannel}.
    \item We propose and analyze a neural network-based fingerprinting with CNN-architecture for positioning in urban area. This method requires low-power and low-memory.
    \item We consider a single BS positioning scenario to reduce the system complexity, and avoid the demand of observing multiple BSs using mmWave signals under high path loss scenarios. In addition, wrapping up the neural network-based processing in a single BS enables efficient scalability of the proposed methods for various use cases, including cases with varying BS deployment.

\end{enumerate}

The remainder of the article is as follows. The related works are discussed in Section \RomanNumeralCaps{2}, the preliminaries are provided in Section \RomanNumeralCaps{3}. Our proposed CNN-based fingerprinting utilizing 5G NR CSI is presented in Section \RomanNumeralCaps{4}, the experiment and analysis are provided in Section \RomanNumeralCaps{5}, followed by conclusion and future work in Section \RomanNumeralCaps{6}.

\section{Related Works}
Fingerprints of radio signals for positioning have been frequently used in the literature. Wu et al. in \cite{csi2}, have utilized Wi-Fi-based CSI data collected by an Atheros CSI tool in an indoor environment. They have proposed a supervised Deep Neural Network (DNN) to estimate the position of the UE. Authors proved that their proposed network architecture outperforms its counterpart in terms of being fast in the online phase and saving memory for the storage of the weights and biases of the network. 

In \cite{csi-wifi2}, DeepNar was proposed as a neural network-based fingerprinting method. Authors have used Wi-Fi Round Trip Time (RTT) fingerprints from 7 BSs to localize a UE in an indoor scenario. There, the RTT measurements are first normalized to stay in the range between [0,1] and then fed to the neural network. By utilizing DeepNar, the authors have achieved a minimum mean positioning error of \highl{$91 \ cm$} with the test samples. \highl{This error is the minimum calculated average error in estimation of test points positions using the test CSI samples.} In \cite{loss}, a deep residual sharing learning based system was presented utilizing Wi-Fi CSI data collected in an indoor environment with the Intel 5300 NIC tool. To prepare the input data for the neural network, authors have constructed CSI tensors utilizing CSI amplitude and AoA measurements. Considering the two types of measurements, the authors feed the data to two different channels. Thus, the trained weights and biases are different in the output of the channels and they will be shared at the end of a residual block. Authors have tested their method in the indoor environment of a laboratory and a corridor, and they were able to achieve the minimum mean error of $1.16 \ m$, \highl{which is the least achieved average error among the estimations for the test samples}. In urban environments, Wi-Fi signals are not always available. 
\\
In \cite{klus2}, authors take advantage of the fingerprints including beamformed radio signal strength in a 5G NR network. They have also fused the position estimations with the available GNSS position data to achieve higher accuracy. 
The considered signals are transmitted by seven beamforming-capable BSs over $32$ beams for each. 
The results in their work show that $3.4 \ m$ mean positioning error is achieved by the proposed fingerprinting method. This error decreased to $1.75 \ m$ mean by fusing the radio signal strength based estimates with GNSS position data utilizing a neural network. In \cite{5g-deep}, authors exploit deep learning-based fingerprinting to localize a UE in an urban area. They have used the Reference Signal Received Power (RSRP) of 5G beams and the beams ID as the input data to their proposed neural network. For simulating the data, the authors have used WinProp ray-tracing simulation tool. They have utilized $8$ 5G radio nodes (gNBs) and achieved the minimum average error of $1.4 \ m$. Authors in \cite{5g_csi} have utilized 5G CSI fingerprints and Siamese CNN to find the position of a UE in an indoor area. They have achieved a minimum mean error of $1.17 \ m$ using $4$ BSs. 
In this work, despite the previous works, we are interested in beamformed CSI fingerprints. Furthermore, regarding the lack of Wi-Fi signals in outdoor environment, we take advantage of the beamforming technology of 5G signals to improve the results for an outdoor scenario. Besides, we utilize only one BS to evaluate the performance of the proposed method in an outdoor positioning scenario.
\\
A summary of methods considered in the literature, which utilize neural networks is presented in Table \ref{fp}. Besides the literature references, the environment type, radio signal fingerprint, and the minimum mean error achieved by the method are provided in the table. \highl{The minimum mean error is the lowest error among the averaged errors achieved in position estimation by feeding the test CSI samples to the trained network.}

\section{Preliminaries with data description}

In this work, we take advantage of the 5G mmWave CSI fingerprints of a MIMO Orthogonal Frequency Devision Multiplexing (OFDM) system to train a CNN. The trained network will be then used for fingerprinting.

\subsection{Neural network based fingerprinting}
In traditional fingerprinting methods, a large amount of data is required during the real-time process to estimate the position of the UE. However, utilizing neural networks enables the system to eliminate the required memory and expense to carry the heavy database in the online phase. 

There are mainly two phases in the neural network-based fingerprinting method: 1) Offline phase: a large amount of CSI data are given as input to the system. The data are labeled with the position of each training point. The weights and biases of the neural network will be optimized based on the given labeled data. Finally, the trained network will be used in the online phase.
2) Online Phase: There will be no requirement of memorizing the training data set. The only necessity will be the weights and biases of the trained network. The test data which is not previously seen by the network will be provided as the input to the network. The network will then estimate the position of the UE.

\subsection{Channel State Information (CSI)}
While the radio signals propagate through the air, they experience different types of channel effects, such as diffractions, specular reflections, scattering, and path loss. The radio channel can be represented by CSI which is essentially a complex-valued frequency response of the channel in OFDM-based transmission. Assuming a single antenna UE, the received signal at the $m^\text{th}$ subcarrier of the $b^\text{th}$ beam can be represented as

\begin{equation}\label{csi1}
r_{m,b} = \mathbf{H}_{m,b} \mathbf{F}_b s_{m,b} + n_{m,b}, 
\end{equation}
\\
where $s_{m,b}$ is the transmitted signal, $\mathbf{H}_{m,b}\in\mathbb{R}^{1 \times N_{\text{TX}}}$ is the channel matrix between the transmitter and the receiver including the effect of all multipaths, $\mathbf{F}_{b}\in\mathbb{R}^{N_{\text{TX}} \times 1}$ is the beamformer vector, and $n_{m,b} \sim \mathcal{N}(0,\sigma^{2})$ is Additive White Gaussian Noise (AWGN). From (\ref{csi1}), the effective beamformed CSI for the $m^\text{th}$ subcarrier and $b^\text{th}$ beam can be observed as $H^{\text{CSI}}_{m,b} = \mathbf{H}_{m,b} \mathbf{F}_b$.  In practice, the CSI is estimated from the received signal assuming a known transmitted reference signal \cite{csi2}. To prepare the input data for the neural network, the available CSI data are received by $M$ number of subcarriers and $B$ number of beams. Considering that the CSI data involve complex numbers, to prepare the CSI data in a way that they are applicable to CNN, we have added the real and imaginary parts of the CSI data separately in two consecutive columns. Using this method prevents any probable data loss during the CNN classification \cite{complex}. Thus, finally, we have the data of $M \times 2B$ matrices as the input to the network.

Assuming a fixed and static environment, separate CSI measurements collected at the same location differ only by measurement noise. In addition, the noise on the CSI data can be considered to vary according to changes in the environment, such as the movement of vehicles \cite{weather}. Thus, for each considered measurement location simulated in the ray-tracing tool, a large amount of data can be generated by obtaining different random noise realizations. The quality of the beam-wise CSI measurements is dependent on the signal SNR, which is defined as
\begin{equation}
     SNR_{b} = 10\log_{10}\left(\frac{\sum_m \vert r_{m,b}\vert^2}{M\sigma^2}\right) \, \text{[dB]},
\end{equation}
where $\sigma^2$ is the noise power.


\subsection{CNN and data splitting}

CNNs have the ability to extract the features of the given input and detect the important features in comparison with other traditional Machine Learning (ML) methods including random forests. The features extraction is performed by an element-wise product between the inputs and a small array of numbers, called a kernel. The kernel is applied to the input by sliding over all the locations to extract the features. Notice that the convolutional layer is linear. Thus, a nonlinear function such as a Rectifying Linear Unit (ReLU) is required to enable the back-propagation. 

Before feeding the data set as the input to the neural network, the data set is first split into the following categories:

\begin{enumerate}
    \item Training data set: Some measurement points on the simulation area are considered as the reference points. A large amount of CSI data are collected at each reference point. Notice that the labels of the CSI data are the positions of each reference point.
    \item Validation data set: This data set is generated for each reference point. However, the validation data has been never seen previously by the network as it has different noise from the training data set. This data set is used to evaluate the robustness of the neural network.
    \item Test data set: In this data set, new points other than reference points are considered to evaluate the network performance and positioning accuracy. For this data set, there will not be any specific labels. Instead of finding only one label, the network will find the probabilities of the labels. The position will then be estimated based on the predicted probabilities. 
\end{enumerate}

These data sets are then fed to the proposed CNN when considering the numerical evaluations shown in Section V.

\section{Proposed CNN-Based Fingerprinting utilizing 5G NR CSI}

\begin{figure}[t]
\centerline{\includegraphics[width=3.2in]{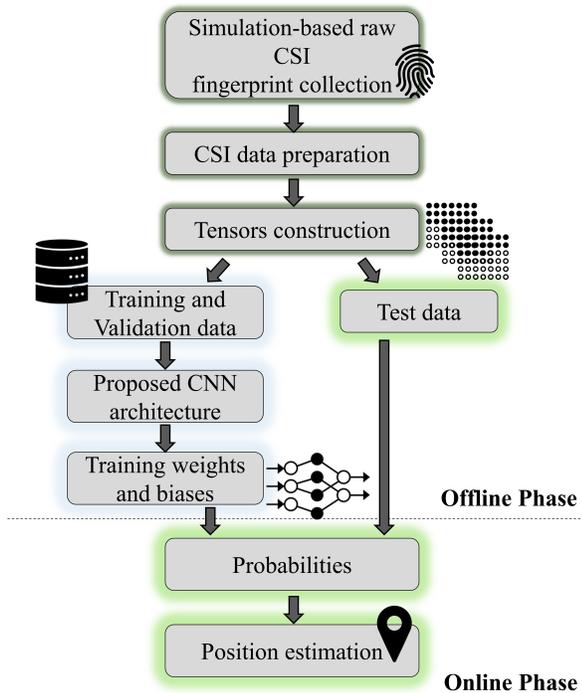}}
  \caption{System architecture of the proposed positioning method.}
  \label{architecture}
\end{figure}

The proposed system architecture is illustrated in Fig. \ref{architecture}, where the CNN is developed to extract the features in CSI data. The hyperparameter of number of layers is optimized by observing the generalization error while increasing the number of layers. The optimum number of layers is selected based on the lowest \highl{validation} error while avoiding overfitting. We have utilized 5 convolutional2D layers with a ReLU function at each layer to define non-linearity. As the downsampling strategy, 2D max-pooling is used in the first 4 layers \cite{maxpool}. Then, in the 5th layer, flattening is done to prepare the input for the output layer of the network, which is the classifier.

\subsection{Hyperparameters}
To achieve the highest accuracy, the choice of hyperparameter of a network plays an important role. Furthermore, having optimal or close to optimal hyperparameters results in saving time, energy, computation, and money \cite{green}. 
Hyperparameters consist of two types of variables: a) the variables which determine the network structure like the number of hidden layers, and b) the variables which affect the training of the network, such as the learning rate, weight decay, and batch size. Moreover, the tuned hyperparameters for the developed CNN are listed in Table \ref{hp}.

Regularization is one of the most essential hyperparameters. In this work, the techniques utilized for regularization are L2 norm and Batch Normalization (BatchNorm). L2 norm is applied with the weight decay and BatchNorm normalizes the input of the layer. It subtracts the input by mean value of the mini-batch and finally divides it by the mini-batch standard deviation. This technique solves the problem of internal covariate shift. This problem occurs due to the varying distribution of the input weights to the neurons at each epoch. Neurons must adapt to this change, which is solved by the BatchNorm technique \cite{covariate}. Another issue that DNNs might experience, is the vanishing or exploding gradient. Vanishing gradients occur due to the gradients getting smaller while the network is being trained. 
On the other hand, exploding gradients occur when big errors accumulate and represent large values to the network. This issue prevents the network to be trained and results in an unstable network. BatchNorm prevents the values to become very big or very small. In this work, we have used a 2D BatchNorm in the first main layer and a 1D BatchNorm in the last main layer after the output has been flattened to a 1D feature vector.  

\begin{table}

\caption{Hyperparameters for Training the Network}
\label{hp}
\begin{center}
    \normalsize
 \begin{tabular}{|c|c|} 
 \hline
\textbf{Hyperparameter} & \textbf{Value} \\ 
 \hline
Activation Function & ReLU\\
\hline
Regularization & BatchNorm and Weight Decay\\
\hline
Decay Rate & 0.001\\
 \hline
Learning Rate & 0.000001\\
\hline
Epochs Size & 150\\
\hline
Batch Size & 20\\
\hline
\end{tabular}
\end{center}
\end{table}

\subsection{Loss Function}
The parameters of the neural network are optimized using AdamW optimizer explained in detail in \cite{adamw}. AdamW is an adaptive optimizer which provides the optimization of weight decay and learning rate separately. The loss function is used to calculate the error in predictions while training the network. We have used two different error calculations during the training and the test phase. 

\subsubsection{Training and the robustness of test performance} For training the network, we aim to minimize the Negative Log Likelihood (NLL) loss function given as
    \begin{equation}\label{nll}
    loss = - \sum_{n=1}^{N}y_{n} \log \hat{y}_{n},
    \end{equation}
    where $N$ is the number of labels (number of reference points in our work), $y_{n}$ is the true label of the $n^\text{th}$ training data and $\hat{y}_{n}$ represent the probabilities computed by the softmax layer as
    \begin{equation}\label{nll}
    \hat{y}_{n} = Softmax(z_{j}) = \frac{\exp{z_{j}}}{\sum_{k=1}^{K}\exp{z_{k}} }, 
    \end{equation}
    where $z_j$ represents the neurons values in output layer. The values are calculated by the input to the neurons, weights and biases. We utilize NLL loss for both training and validation. The validation is done by CSI data which have never been seen by the network previously. However, this CSI data have the same labels with the reference points.
    
    \subsubsection{New Test Points} The final test is done for the CSI data collected at the points other than reference points. \highl{The loss/error in the predictions which are estimated positions for the test points is the Euclidean distance to the ground truth.}
    
    
 The predicted coordinates \highl{($x_{pred}$,$y_{pred}$)} for the corresponding CSI data are estimated using the softmax layer probabilities as
    
    
    \begin{align}\label{x_predict}
    \begin{split}
    x_{pred} &= \sum_{i=1}^{R} \frac{p_{i} }{\sum_{j=1}^{R}p_{j}}  x_{i} \text{  and} \\
    y_{pred} &= \sum_{i=1}^{R} \frac{p_{i} }{\sum_{j=1}^{R}p_{j}} y_{i},
    \end{split}
    \end{align}
where $(x_{i},y_{i})$ represents the position of the training points, $p_i$ is the probability calculated by the softmax layer, and $R$ is the number of selected training points, defined as $R=4$ in this work. The $R$ selected training points have the largest calculated probabilities $p_i$ among the others.
\highl{The estimator is sensitive to the value of $R$. $R=4$ is selected considering the consistency it provides in the accuracy of calculated positions. Higher values of R result in lower accuracy and lower values of R provides inconsistent solutions.}

\begin{figure}
  \centerline{\includegraphics[width=3.5in]{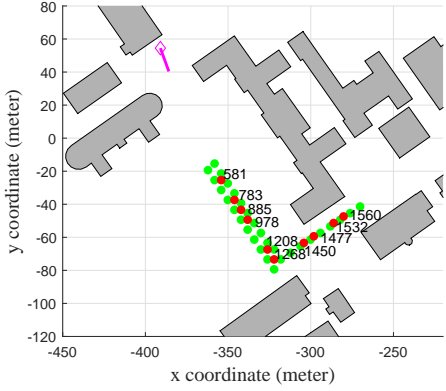}}
  \caption{Reference and test points in the simulated environment. The green circles are the reference points, and the red circles with unique label numbering are the test points. The BS location is shown with a magenta marker accompanied with a pointer vector to illustrate the antenna direction.}
  \label{map}
\end{figure}


\section{Numerical experiment and Analysis}

The proposed method is tested by simulating CSI data in an urban environment in Helsinki metropolitan area called Punavuori. In order to obtain the CSI data, we utilize a proprietary ray-tracing software, following the map-based channel modeling principles proposed by the 3GPP in \cite{3GPPchannel}. In total of $30$ number of training point locations and $10$ number of test point locations are considered in the environment, as shown in Fig. \ref{map}. The carrier frequency is defined as $30$~GHz.

At each location, the CSI is measured from a single BS based on a downlink transmitted Synchronization Signal Blocks (SSBs)  consisting of 240 subcarriers with subcarrier spacing of $60$~kHz (bandwidth of $14.4$~MHz). Furthermore, the SSBs are beamformed over $B=32$ beams using a uniform rectangular array with dimensions of 16 azimuth and 8 elevation elements. For each reference point, $1000$ noisy CSI samples are obtained for training and validation, as discussed in Section III.B. The radio link budget is designed so that in Line-of-Sight (LOS) conditions, the SNR for the best beam is approximately $10$~dB at $100$~m distance from the BS. From the generated data 60\% is used  for training and 40\% for validation \highl{using random splitting}. Similarly, for testing purposes, $100$ CSI samples are correspondingly generated for the test point locations.


\begin{figure}[t]
  \centerline{\includegraphics[width=3.2in]{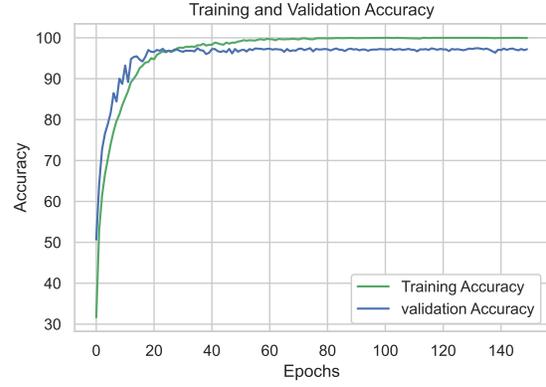}}
  \caption{The accuracy of training and validation.}
  \label{acc}
\end{figure}

The training and validation accuracy are presented in Fig.~\ref{acc}. Considering the training accuracy, we can realize that the network has been appropriately trained. It shows that the proposed CNN is able to extract the features out of the CSI tensors. 
Furthermore, the validation accuracy, extracted from the validation set, has reached maximum $97.77 \%$. 
The high validation accuracy indicates that first the network is well-regularized due to regularization methods, and second, the network is robust against the unseen data.

\begin{figure}[t]
  \centerline{\includegraphics[width=3.2in]{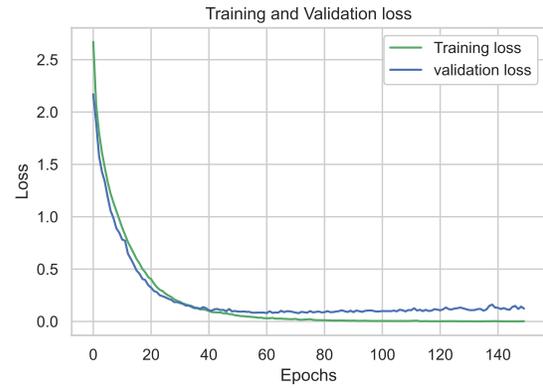}}
  \caption{The loss of training and validation.}
  \label{loss}
\end{figure}

The training and validation loss are presented in Fig.~\ref{loss}. It is shown that the training loss values are decreasing to a point of stability and the optimized model fits well to the training data. 
Moreover, considering that also the validation loss is decreasing to a good extent, the model seems to perform adequately. However, it can be seen that the validation loss begins to slightly increase around the $80^\text{th}$ epoch. \highl{This can be a result of the data shuffling at each epoch. It could be the case that the random sampler has taken a more complex data at these epochs in comparison with the previous epochs. However, at this point} we expect the regularization method to modify the learning algorithm so that the model generalizes in the optimum way. 

\begin{figure}[t]
  \centerline{\includegraphics[width=3.2in]{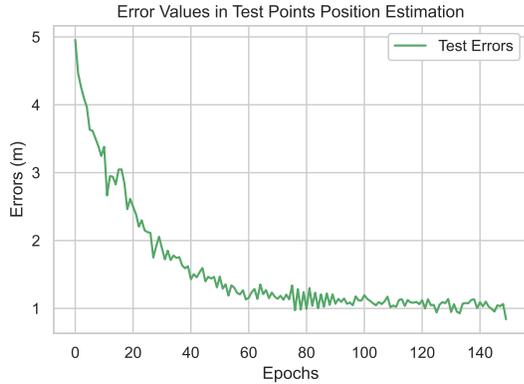}}
  \caption{The mean absolute error of positioning for the test points.}
  \label{errors}
\end{figure}

The mean absolute positioning error for the test point locations is presented in Fig.~\ref{errors}. \highl{The time required for our system to estimate the position of a test point by feeding one CSI sample to the network is $8 \ ms$.} It should be yet emphasized that the test data is collected at different locations compared to the training data, and thus, by enhancing the generalization of the network, it should be possible to reduce the positioning error. Although the network is able to consistently reduce the positioning error when increasing the number of epochs, it is seen that similarly as with the validation loss, the network struggles around the $80^\text{th}$ epoch. \highl{It seems that it took a few epochs for the regularization method to modify the learning algorithm and reduce the generalization error and achieve the consistent downward trend in test error.}  

\begin{figure}[h]
  \centerline{\includegraphics[width=3.2in]{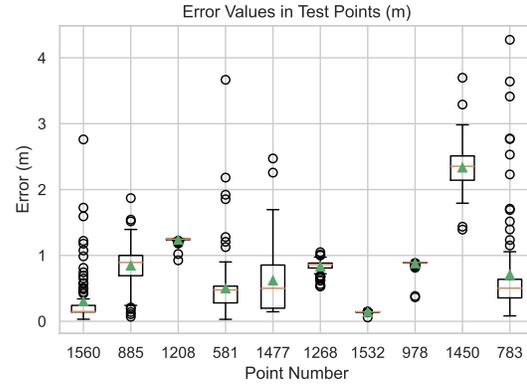}}
  \caption{\highl{The distribution of positioning errors at each test point.}}
  \label{dist}
\end{figure}

In order to analyze the positioning errors at different map locations, the error distribution at each test point is illustrated in Fig. \ref{dist}. It is shown that in most of the points, the average error is below one meter. \highl{Considering that points number $1532$ and $1560$ are in NLOS conditions and the rest are in LOS conditions, one noticeable observation is that the proposed method works accurately also in NLOS conditions}. From the CSI perspective this is justified by the fact that the CSI is more unique per location in NLOS conditions compared to LOS-dominated scenarios, where nearby locations share similar multipath characteristics with only a slight power difference. Thus, on contrary to traditional positioning methods, machine learning based methods seem to naturally manage positioning in both LOS and NLOS conditions in the considered urban environment. 

\section{Conclusion and Future Work}
In this work, we have presented a deep learning-based position estimation method using 5G CSI fingerprints and beamforming technology for the first time in the literature. We have taken advantage of the neural networks to find the static single BS-based position of a UE in a specific outdoor area. The CSI fingerprints of 5G NR mmWave have been collected based on ray-tracing simulations. The results show that by using machine learning-based CSI fingerprinting with CNNs, it is possible to achieve a sub-meter positioning error regardless of the LOS condition of the channel.

In the next steps of this research, \highl{ higher number of reference points in a wider area will be considered. Furthermore, }a mobile user positioning will be investigated by utilizing the time series and recurrent neural networks with measurements from several BSs.


\bibliography{ref}

\begin{thebibliography}{10}
\providecommand{\url}[1]{#1}
\csname url@samestyle\endcsname
\providecommand{\newblock}{\relax}
\providecommand{\bibinfo}[2]{#2}
\providecommand{\BIBentrySTDinterwordspacing}{\spaceskip=0pt\relax}
\providecommand{\BIBentryALTinterwordstretchfactor}{4}
\providecommand{\BIBentryALTinterwordspacing}{\spaceskip=\fontdimen2\font plus
\BIBentryALTinterwordstretchfactor\fontdimen3\font minus
  \fontdimen4\font\relax}
\providecommand{\BIBforeignlanguage}[2]{{%
\expandafter\ifx\csname l@#1\endcsname\relax
\typeout{** WARNING: IEEEtran.bst: No hyphenation pattern has been}%
\typeout{** loaded for the language `#1'. Using the pattern for}%
\typeout{** the default language instead.}%
\else
\language=\csname l@#1\endcsname
\fi
#2}}
\providecommand{\BIBdecl}{\relax}
\BIBdecl

\bibitem{klus2}
R.~Klus, J.~Talvitie, and M.~Valkama, ``{Neural Network Fingerprinting and GNSS
  Data Fusion for Improved Localization in 5G},'' in \emph{2021 International
  Conference on Localization and GNSS (ICL-GNSS)}, 2021, pp. 1--6.

\bibitem{5g-deep}
M.~M. Butt, A.~Rao, and D.~Yoon, ``{RF Fingerprinting and Deep Learning
  Assisted UE Positioning in 5G},'' in \emph{2020 IEEE 91st Vehicular
  Technology Conference (VTC2020-Spring)}, 2020, pp. 1--7.

\bibitem{rets}
G.~Retscher, V.~Gikas, H.~Hofer, H.~Perakis, and A.~Kealy, ``{Range Validation
  of UWB and Wi-Fi for Integrated Indoor Positioning},'' \emph{Applied
  Geomatics}, vol.~11, 01 2019.

\bibitem{aoa}
A.~Shahmansoori, B.~Uguen, G.~Destino, G.~Seco-Granados, and H.~Wymeersch,
  ``{Tracking Position and Orientation Through Millimeter Wave Lens MIMO in 5G
  Systems},'' \emph{IEEE Signal Processing Letters}, vol.~26, no.~8, pp.
  1222--1226, 2019.

\bibitem{csi2}
G.-S. Wu and P.-H. Tseng, ``{A Deep Neural Network-Based Indoor Positioning
  Method using Channel State Information},'' in \emph{2018 International
  Conference on Computing, Networking and Communications (ICNC)}, 2018, pp.
  290--294.

\bibitem{loss}
X.~Wang, X.~Wang, and S.~Mao, ``{Indoor Fingerprinting With Bimodal CSI
  Tensors: A Deep Residual Sharing Learning Approach},'' \emph{IEEE Internet of
  Things Journal}, vol.~8, no.~6, pp. 4498--4513, 2021.

\bibitem{g-book}
P.~Groves, \emph{{Principles of GNSS, inertial, and multisensor integrated
  navigation systems, Second edition}}, 03 2013.

\bibitem{ION}
N.~Dvorecki, O.~Bar-Shalom, L.~Banin, and Y.~Amizur, ``{A machine learning
  approach for Wi-Fi RTT ranging},'' in \emph{In Proceedings of the 2019
  International Technical Meeting of The Institute of Navigation}, 01 2019, pp.
  435--444.

\bibitem{csi-wifi}
X.~Tong, Y.~Wan, Q.~Li, X.~Tian, and X.~Wang, ``{CSI Fingerprinting
  Localization With Low Human Efforts},'' \emph{IEEE/ACM Transactions on
  Networking}, vol.~29, no.~1, pp. 372--385, 2021.

\bibitem{wifi_fp}
D.~Quezada-Gaibor, J.~Torres-Sospedra, J.~Nurmi, Y.~Koucheryavy, and J.~Huerta,
  ``{Lightweight Wi-Fi Fingerprinting with a Novel RSS Clustering Algorithm},''
  in \emph{2021 International Conference on Indoor Positioning and Indoor
  Navigation (IPIN)}, 2021, pp. 1--8.

\bibitem{complex}
A.~Vora, P.-X. Thomas, R.~Chen, and K.-D. Kang, ``{CSI Classification for 5G
  via Deep Learning},'' in \emph{2019 IEEE 90th Vehicular Technology Conference
  (VTC2019-Fall)}, 2019, pp. 1--5.

\bibitem{wym1}
H.~Wymeersch, G.~Seco-Granados, G.~Destino, D.~Dardari, and F.~Tufvesson, ``{5G
  mmWave Positioning for Vehicular Networks},'' \emph{IEEE Wireless
  Communications}, vol.~24, no.~6, pp. 80--86, 2017.

\bibitem{beamforming}
F.~{Sohrabi} and W.~{Yu}, ``{Hybrid Digital and Analog Beamforming Design for
  Large-Scale Antenna Arrays},'' \emph{IEEE Journal of Selected Topics in
  Signal Processing}, vol.~10, no.~3, pp. 501--513, 2016.

\bibitem{csi-wifi2}
O.~Hashem, K.~A. Harras, and M.~Youssef, ``{DeepNar: Robust Time-based
  Sub-meter Indoor Localization using Deep Learning},'' in \emph{2020 17th
  Annual IEEE International Conference on Sensing, Communication, and
  Networking (SECON)}, 2020, pp. 1--9.

\bibitem{5g_csi}
Q.~Li, X.~Liao, M.~Liu, and S.~Valaee, ``{Indoor Localization Based on CSI
  Fingerprint by Siamese Convolution Neural Network},'' \emph{IEEE Transactions
  on Vehicular Technology}, vol.~70, no.~11, pp. 12\,168--12\,173, 2021.

\bibitem{3GPPchannel}
3GPP, ``{Study on channel model for frequencies from 0.5 to 100 GHz positioning
  in NG-RAN},'' Tech. Rep. 38.901, 12 2019, version 16.1.0.

\bibitem{weather}
C.~Luo, J.~Ji, Q.~Wang, X.~Chen, and P.~Li, ``{Channel State Information
  Prediction for 5G Wireless Communications: A Deep Learning Approach},''
  \emph{IEEE Transactions on Network Science and Engineering}, vol.~7, no.~1,
  pp. 227--236, 2020.

\bibitem{maxpool}
B.~Zhao, Y.~S. Chong, and A.~Tuan~Do, ``{Area and Energy Efficient 2D
  Max-Pooling For Convolutional Neural Network Hardware Accelerator},'' in
  \emph{IECON 2020 The 46th Annual Conference of the IEEE Industrial
  Electronics Society}, 2020, pp. 423--427.

\bibitem{green}
E.~Strubell, A.~Ganesh, and A.~Mccallum, ``{Energy and Policy Considerations
  for Deep Learning in NLP},'' in \emph{Proceedings of the 57th Annual Meeting
  of the Association for Computational Linguistics}.\hskip 1em plus 0.5em minus
  0.4em\relax Association for Computational Linguistics, 01 2019, pp.
  3645--3650.

\bibitem{covariate}
\BIBentryALTinterwordspacing
S.~Ioffe and C.~Szegedy, ``{Batch Normalization: Accelerating Deep Network
  Training by Reducing Internal Covariate Shift},'' \emph{CoRR}, vol.
  abs/1502.03167, 2015. [Online]. Available:
  \url{http://arxiv.org/abs/1502.03167}
\BIBentrySTDinterwordspacing

\bibitem{adamw}
R.~Llugsi, S.~E. Yacoubi, A.~Fontaine, and P.~Lupera, ``{Comparison between
  Adam, AdaMax and AdamW Optimizers to Implement a Weather Forecast based on
  Neural Networks for the Andean City of Quito},'' in \emph{2021 IEEE Fifth
  Ecuador Technical Chapters Meeting (ETCM)}, 2021, pp. 1--6.

\end{thebibliography}
\bibliographystyle{IEEEtran}

\end{document}